\documentclass[conference]{IEEEtran}
\usepackage{graphicx}
\usepackage{listings}
\usepackage{courier}
\usepackage{cite}
\usepackage[]{footmisc}
\usepackage{url}

\makeatletter
\@ifundefined{showcaptionsetup}{}{%
 \PassOptionsToPackage{caption=false}{subfig}}
\usepackage{subfig}
\makeatother

\def\BibTeX{{\rm B\kern-.05em{\sc i\kern-.025em b}\kern-.08emT\kern-.1667em\lower.7ex\hbox{E}\kern-.125emX}}

\begin{document}

\title{Manticore: A User-Friendly Symbolic Execution Framework for Binaries and Smart Contracts}

\author{\IEEEauthorblockN{Mark Mossberg, Felipe Manzano, Eric Hennenfent, Alex Groce, \\
Gustavo Grieco, Josselin Feist, Trent Brunson, Artem Dinaburg}\\
\IEEEauthorblockA{Trail of Bits, New York City, USA\\
Email: \{mark, felipe, eric.hennenfent, alex.groce, gustavo.grieco, josselin, trent.brunson, artem\}@trailofbits.com}
}

%


\newcommand{\jf}[1]{{\color{blue}#1}}
\lstset{
columns=flexible,breaklines=true,breakatwhitespace=true,basicstyle=\small\ttfamily}

%
%


%



%
\maketitle

\begin{abstract}
An effective way to maximize code coverage in software tests is through dynamic symbolic execution---a technique that uses constraint solving to systematically explore a program's state space. We introduce an open-source dynamic symbolic execution framework called Manticore for analyzing binaries and Ethereum smart contracts. Manticore's flexible architecture allows it to support both traditional and exotic execution environments, and its API allows users to customize their analysis. Here, we discuss Manticore's architecture and demonstrate the capabilities we have used to find bugs and verify the correctness of code for our commercial clients.

\end{abstract}

\section{Introduction}
%

Dynamic symbolic execution is a program analysis technique that explores a state space with a high degree of semantic awareness~\cite{Boyer:1975}.
For paths that are explored by the analysis, dynamic symbolic execution identifies a set of \emph{path predicates}: constraints on the program's input. These are used to generate program inputs that will cause the associated paths to execute.
This approach produces no false positives in the sense that all identified program states can be triggered during concrete execution. For example, if the analysis finds a memory safety violation, it is guaranteed to be reproducible.

Symbolic execution has been extensively researched in a security context~\cite{Baldoni:2018}, but industry has been slow to adopt the technique because of the limited availability of flexible, user-friendly, tools. Furthermore, existing frameworks are tightly coupled to traditional execution models, which makes symbolic execution research challenging for alternative execution environments, such as the Ethereum platform. 

Manticore is a symbolic execution framework for analyzing binaries and smart contracts.
Trail of Bits has used this tool internally in numerous code assessments~\cite{Manticore-Paxos, Manticore-Golem, Manticore-Sai, Manticore-Gemini, Manticore-Ampleforth}, and in program analysis research, including the DARPA Cyber Grand Challenge~\cite{cgc} (CGC).  



\section{Architecture}

Manticore's design is highly flexible and supports both traditional computing environments (x86/64, ARM) and exotic ones, such as the Ethereum platform. To our knowledge, it is the only symbolic execution framework that caters to such different environments. It is also simple, extensible, and as self-contained as possible, avoiding unwarranted external dependencies.

\begin{figure*}
\subfloat[A high-level architecture diagram. \label{subfig:design}]{\includegraphics[width=0.60\textwidth]{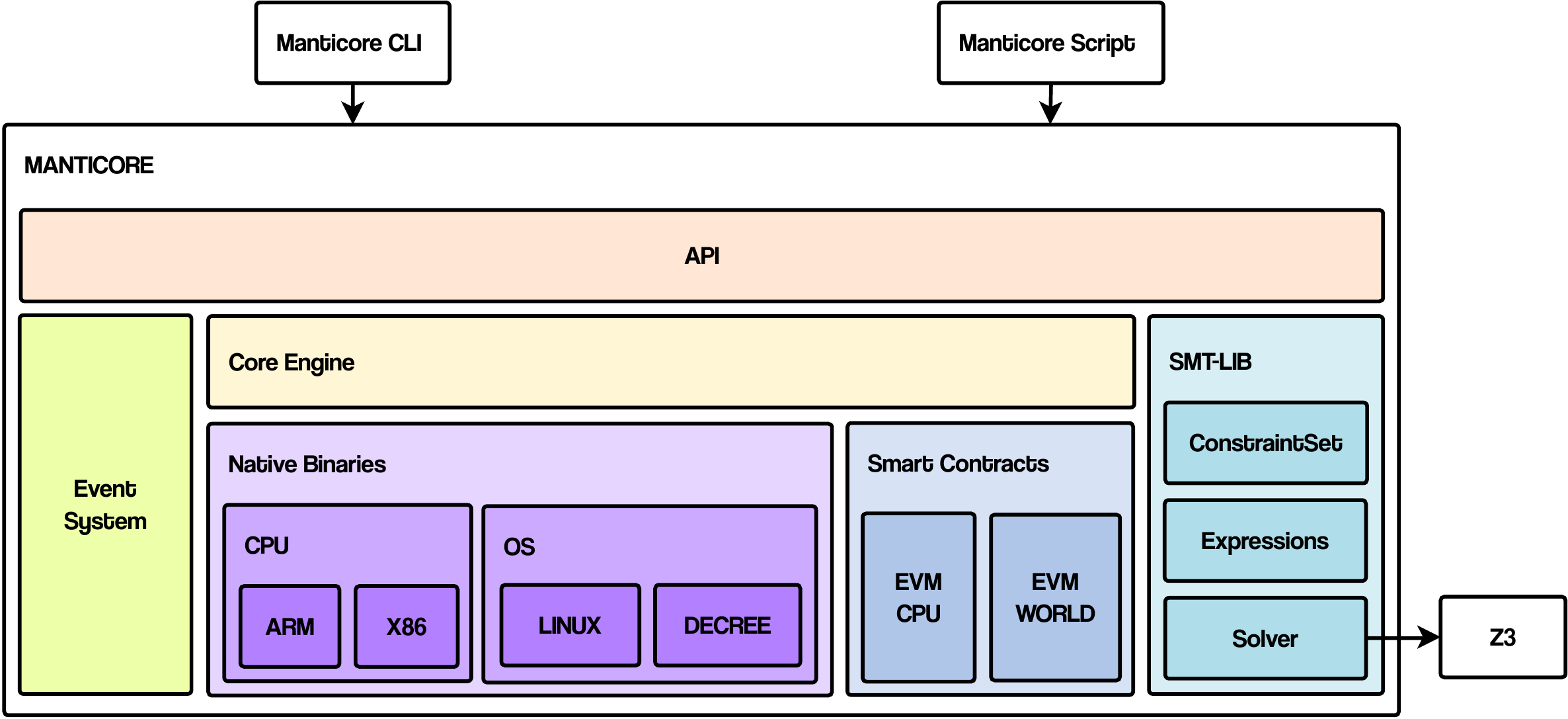}

}\subfloat[The state life cycle. \label{subfig:lifecycle}]{\includegraphics[width=0.30\textwidth]{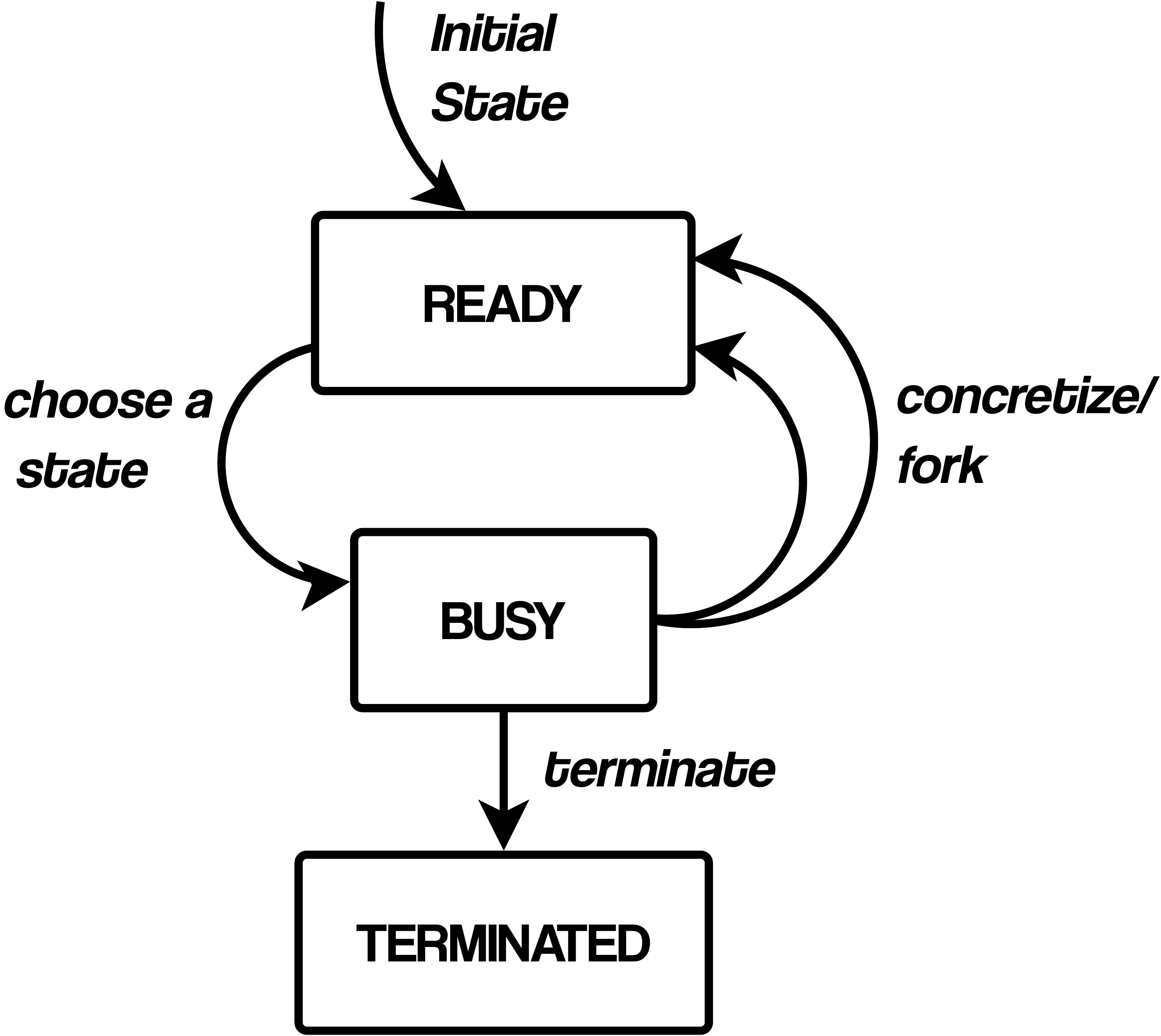}

}

\caption{Manticore Overview}

\end{figure*}

Figure~\ref{subfig:design} shows Manticore's architecture. The primary components are the Core Engine and Native and Ethereum Execution Modules. Secondary components include the Satisfiability Modulo Theories (SMT-LIB) module, Event System, and API.



\subsection{Core Engine}
The Core Engine is the source of Manticore's flexibility. It implements a generic platform-agnostic symbolic execution engine that makes few assumptions about the underlying execution model. 

This Core Engine operates and manages program states according to the State Life Cycle shown in Figure \ref{subfig:design}.
Program states are abstract objects that represent the state of a program at a point in execution.
These objects expose an execution interface that the Core Engine invokes to trigger one atomic unit of program execution. For native binaries and Ethereum, this is one instruction. During execution, states can interrupt back to the Core Engine to signal that a life cycle event needs to be handled.

The State Life Cycle, shown in Figure \ref{subfig:lifecycle}, defines three states: \emph{Ready}, \emph{Busy}, and \emph{Terminated} and two events: \emph{Termination} and \emph{Concretization}. 
The Core Engine repeatedly selects a \emph{Ready} state and executes it (transitioning it to \emph{Busy}).
An executing \emph{Busy} state can either transition back to \emph{Ready} or signal a Life Cycle event for the Core to handle. 
 
The \emph{Termination} event occurs when a state reaches an end, typically on program exit or a memory access violation, which transitions the state to \emph{Terminated}.
\emph{Concretization} happens when a state signals that a symbolic object should be converted into one or more concrete values, subject to the current constraints on the State. 
For each concrete value, one new child State is created and marked \emph{Ready}. The most common case of \emph{Concretization}, called forking, occurs when a program counter register becomes symbolic and is concretized to possible concrete values. This causes new states to be generated for each new program path. 

State exploration can be customized using various policies, which implement a variety of heuristics for \emph{Ready} state selection and \emph{Concretization}. The Core Engine was designed for parallelism and supports multiple processes for state queue processing.
\subsection{Native Execution Module}
The native binary symbolic execution module abstracts hardware execution to implement the high-level execution interface that the Core Engine expects, via symbolic emulation of the CPU, memory, and operating system interfaces.
Currently, the native execution module emulates Linux on x86, x86\_64, ARMv7, and AArch64 as well as DECREE~\cite{cgc} on x86.  

\subsubsection{CPU Emulation}

The symbolic CPU emulation is straightforward and follows the ISA specification directly, with no intermediate representation. Emulated registers and instructions must support both concrete and symbolic values, which they do by building a symbolic expression tree, as opposed to performing computation directly.




%

\subsubsection{Memory Emulation}

Manticore has a simple virtual address space emulation with interfaces for reading, writing, and managing memory mappings. Different policies for handling symbolic memory accesses are implemented. These include fully symbolic and concretized memory models.

\subsubsection{Operating System Emulation}
Manticore includes OS support for the Linux and DECREE operating systems, emulating the system call (syscall) interface, interfaces related to a process address space (e.g. auxiliary vectors, thread local storage), and miscellaneous state setup (e.g. binary loading).
Syscalls must handle symbolic inputs, yet few can be reasonably modeled symbolically. Manticore therefore concretizes system call arguments and (much like KLEE~\cite{klee}) forwards such calls to the real OS.

\subsection{Ethereum Execution Module}
Manticore supports Ethereum \emph{smart contracts}, which are applications compiled according to the Ethereum Virtual Machine (EVM) specification that run on the Ethereum blockchain.
Smart contracts are essentially state machines, commonly used to implement financial instruments, such as auctions and wallets for custom currencies~\cite{smart-contracts-applications}.
There are many differences between EVM and traditional execution. A few examples include a "gas" cost for executing instructions, radically different memory and persistent storage models, and execution state rollbacks.
Despite these differences, adding Ethereum support did not require substantial architectural changes to Manticore, since the Core Engine is completely decoupled from all execution platform details.

\subsubsection{Ethereum Symbolic Execution}

Smart contracts receive input as network transactions consisting of a \emph{value} and a data buffer. The transaction data buffer contains information about which function should be executed in a contract, and its arguments.

Symbolic execution of smart contracts involves \emph{symbolic transactions}, where both value and data are symbolic. Symbolic transactions are applied to all \emph{Ready} states, which cause the symbolic execution of one transaction.
Symbolic transactions can be repeatedly executed to \emph{generically} explore the state space of a contract.

Manticore's emulated environment for smart contract execution supports an arbitrary number of interacting contracts. It is capable of tracking not only a single contract's state, but a full Ethereum "world", with multiple interacting contracts. Manticore has support for handling symbolic indexing, and can even support the SHA3 EVM instruction (despite the inherent difficulty in symbolically executing hash functions) using techniques derived from \cite{godefroid2011higher}.



%



\subsection{Auxiliary Modules}
Manticore also has auxiliary modules like the SMT-LIB module that supplies a custom symbolic expression object model and an SMT solver interface. Different solvers can be used seamlessly, since Manticore interacts with solvers via the SMT-LIB language.

The Event System decouples Manticore as a whole from external instrumentation-based analyses. Arbitrary points within Manticore can broadcast various symbolic execution events (e.g. memory reads/writes, state forking, concretization) that can be handled by an event subscriber outside of Manticore, such as an API client. This provides the foundation for Manticore's plugin system allowing users to create modular, event-based analyses.

\section{Usage}

Manticore has a command-line interface and an API that works for both binaries and smart contracts. An example command follows,

\begin{lstlisting}
$ manticore target ++ +++.txt --data AA --procs 10
\end{lstlisting}

+The arguments \texttt{target ++ +++.txt} instruct Manticore to execute \texttt{target} with two arguments. Manticore uses the "+" character as a stand-in for a symbolic byte, so the first argument is a 2-byte string of symbolic data. The second is a mixed symbolic/concrete string with five bytes of symbolic data followed by the concrete bytes \texttt{.txt}. \texttt{--data} specifies concrete bytes to prefix the stdin input stream, which by default contain 256 symbolic bytes. \texttt{--procs} allocates 10 cores to the analysis. Manticore's output is a directory containing generated inputs and information about each discovered state, as shown below.

\vspace{1em}
{\small{}
\texttt{
\hspace{-2.5em}
\begin{tabular}{@{}ll}
    \$ ls mcore\_x2gncpcq/ &\\
    test\_00000000.argv  &    test\_00000000.input\\
    test\_00000000.messages & test\_00000000.smt\\
    test\_00000000.stdin  &    test\_00000000.trace\\
    test\_00000000.stdout &   ...
 \end{tabular}
}
}
\vspace{1em}

For example, \verb|test_00000000.stdin| can be piped directly to the stdin of the program during concrete execution to trigger the execution state corresponding to \verb|test_00000000|.

Manticore's Python API allows advanced users to customize their analysis using various forms of instrumentation. 
Hooking via the API lets users execute callbacks when a certain state is reached. The callback can access the corresponding State object, which allows complete control over the emulated state. CPU registers, memory, and operating system state can be read, written, filled with symbolic bytes, or concretized. Moreover, states can be pruned, custom constraints can be applied, and satisfiability queries can be sent to the solver. Writing code using the hook API is relatively straightforward, e.g.:

\begin{lstlisting}[label={lst:hooking}, language=Python]
from manticore.native import Manticore
m = Manticore.linux('./target')
@m.hook(0x400ca0)
def hook(state):
    # Disregard state if RDX can be equal 0x44
    # (RDX could be symbolic)
    if state.can_be_true(state.cpu.RDX == 0x44)
        state.abandon()
    input_buf = state.new_symbolic_buffer(32)
    # Apply arbitrary preconstraint on input buffer
    state.constrain(input_buf[0] != ord('A'))
    # Write symbolic buffer at address RBX
    state.cpu.write_bytes(state.cpu.RBX, input_buf)
\end{lstlisting}

Ethereum usage is similar, including a simple command line interface and extensive API for instrumentation.  More details on Manticore's functionality are available online~\cite{ManticoreDocs}.
\section{Native Binary Analysis Evaluation}
\begin{table}[t!]
\centering
\caption{Logic bomb benchmark results (300s)}
\begin{tabular}{|c|c|c|c|c|}
\cline{2-5} 
\multicolumn{1}{c|}{} & Manticore & Angr & Triton & KLEE\tabularnewline
\hline 
Passed & 16 & 17 & 3 & 10\tabularnewline
\hline 
Failed & 33 & 28 & 57 & 37\tabularnewline
\hline 
Timed out & 14 & 18 & 3 & 7\tabularnewline
\hline 
Inapplicable & 0 & 0 & 0 & 9\tabularnewline
\hline 
\end{tabular}
  \label{tab:logic-bombs}
  \vspace{-0.75cm}
\end{table}

We evaluated Manticore's native binary analysis precision and performance using the logic bomb benchmark suite for symbolic execution engines~\cite{benchmarking-with-logic-bombs}.  
This benchmark suite includes a set of 63 logic bombs. Logic bombs are small programs designed to be \emph{triggered} when certain conditions are met. These were designed to test dynamic testing tools, in particular symbolic execution ones.
Logic bombs in this suite are divided into \emph{symbolic-reasoning} and \emph{path-explosion} challenges.
The \emph{symbolic-reasoning} challenges  include paths that potentially produce incorrect test cases. The \emph{path-explosion} challenges are designed to produce too many potential paths, exhausting the resources available for exploration. 

We ran the latest revision of Manticore (revision \texttt{3ffafd5}) on all the logic bomb challenges, and compared the results with angr (8.19.4.5), Triton (0.5), and KLEE (revision \texttt{3ef59a4}). We performed our experiments\footnote{\label{note1}The data and code to reproduce our experiments is available here: \\ \url{https://gist.github.com/ggrieco-tob/76e41835681a59d91fd73b6f5bb7bbe4}} 
using Ubuntu 18.04 on an Intel i7 and 16 GB of RAM, with a $300$ second timeout as suggested by the authors of the benchmark.

Table \ref{tab:logic-bombs} shows the results ({\bf Passed} is the best result possible). Manticore performs almost as effectively as angr, and solved three logic bombs angr was unable to handle. Since angr supports the use of some IEEE 754 floating point instructions that Manticore does not, we expect that Manticore may overtake angr in the future, once it adds such support. If we compare the current results with the ones reported by the original authors of the benchmark~\cite{benchmarking-with-logic-bombs}, the precision of angr has decreased: a few tests are no longer passing. For this benchmark, Triton and KLEE perform relatively poorly. 

Manticore is also fully integrated into the DeepState parameterized unit testing tool~\cite{deepstate}, where it has proved useful in cases where angr failed to produce useful results.

\section{Ethereum Smart Contract Analysis Evaluation}

We evaluated Manticore based on a corpus of 100 Ethereum smart contracts taken directly from the Ethereum blockchain
We ran an analysis that repeatedly executes symbolic transactions against a contract, and tracks the number of states discovered and coverage of the contract code.

\begin{figure}[t!]
  \includegraphics[width=0.40\textwidth]{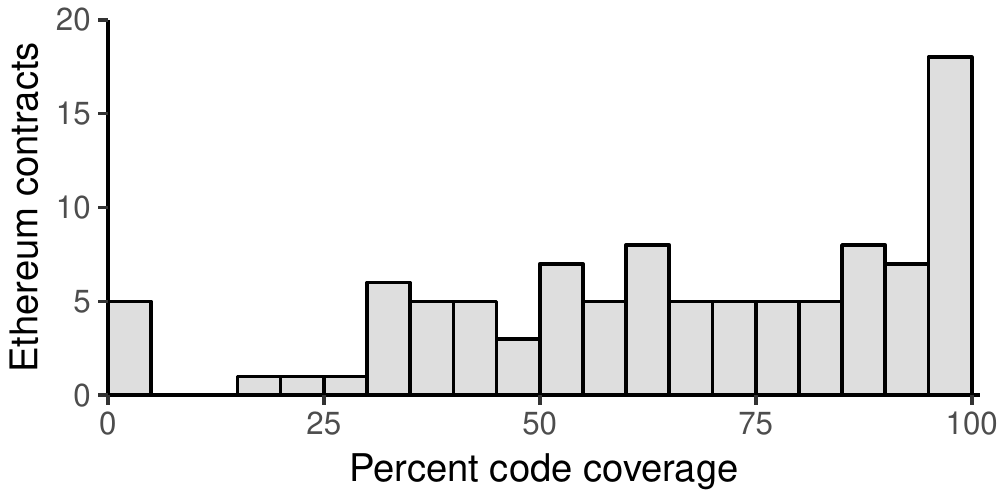}
  \vspace{-0.2cm}
  \caption{Ethereum contract code coverage (n=100)}
  \label{fig:coverage}
\end{figure}

We report the results of running this analysis, with a timeout of 90 minutes per contract, in Table \ref{tab:etheval}.
Manticore produced an average coverage of 65.64\%, with an approximately equal median.  The mean total number of (symbolic) states reached was 207.71, with median 52,  showing that there were a number of outliers where more states were discovered.
Coverage ranged from 0\% to 100\% (see Figure \ref{fig:coverage}).  Four contracts failed in the account creation stage due to incorrect assumptions our analysis script made about the the expected world state, causing them to report 0\% coverage. These assumptions were, however, sufficient for Manticore to achieve greater than 90\% code coverage on 25 of the contracts. 
While useful as a rough evaluation metric, analyzing random smart contracts without context does not mirror our typical process for using Manticore on real-world code-bases. Manticore can achieve much higher code coverage when the initial Ethereum world state is appropriately tailored to the contract's expectations.  The results are nonetheless encouraging, given the lack of assistance.

\begin{table}[t!]
\caption{Ethereum smart contract evaluation results}
\begin{tabular}{l|lllll}
                  & Mean   & Median & Std. Dev. & Min  & Max     \\ \hline
Coverage          & 66\%  & 66\%  & 27\%     & 0\% & 100\%  \\
Running States    & 78  & 16  & 154    & 0 & 998  \\
Terminated States & 130 & 36  & 219    & 1 & 1129 \\
Total States      & 208 & 52  & 356    & 1 & 2127 \\ \hline
\end{tabular}
\label{tab:etheval}
\vspace{-0.75cm}
\end{table}

\subsection{Smart Contract Security Assessments}

Manticore has been used in a number of Trail of Bits client engagements for bug discovery and verification of code invariants. 
Code assessment reports~\cite{Manticore-Paxos, Manticore-Golem, Manticore-Sai, Manticore-Gemini, Manticore-Ampleforth} provide more detail on using Manticore to find bugs or verify code on real-world smart contracts.


\section{Related Work}
The past decade has seen a resurgence in interest in symbolic execution and there are a variety of prominent existing symbolic execution tools. 
Though not strictly a binary analysis tool, KLEE~\cite{klee} was one of the first widely-used symbolic execution implementations. angr~\cite{angr} is a well known binary analysis framework, including extensive symbolic execution functionality. Triton, binsec~\cite{binsec}, and miasm are other well-known binary symbolic execution tools.

Symbolic execution in the Ethereum space is much less widely explored, but other tools do exist, including Mythril~\cite{MythrilClassic}, VerX~\cite{permenevverx}, and KEVM~\cite{kevm}. Like Manticore, all of these tools have been used in commercial software audits to some success. To the best of our knowledge, Manticore is unique among these tools as the only one that was not purpose-built for Ethereum.

\section{Conclusion}

In this paper, we present Manticore, a dynamic symbolic execution framework. Manticore includes user-friendly interfaces and a flexible architecture that allows it to uniquely support execution platforms as diverse as traditional binaries and the Ethereum platform. This is primarily facilitated by the Core Engine, whose symbolic engine logic is decoupled from details of a particular execution environment.
Manticore is useful for leveraging symbolic execution to test programs, and also perform symbolic execution research on alternative execution platforms.
A simple command-line interface is included as well as an API that allows the user to build custom symbolic execution-based tools.
Our evaluation shows that Manticore performs comparably to another standard symbolic execution tools for regular binaries, and on average achieves 66\% code coverage with a default smart contract analysis.
Manticore is open source, and can be found at: \textit{https://github.com/trailofbits/manticore}.

%

%

\bibliographystyle{plain}
\bibliography{bibliography}

\end{document}